# Quark confinement and metric fluctuations

P. R. SILVA

e-mail:prsilvafis@terra.com.br

DEPARTAMENTO DE FÍSICA - ICEx

UNIVERSIDADE FEDERAL DE MINAS GERAIS

CAIXA POSTAL 702

30161-970 - BELO HORIZONTE - MG

BRASIL

ABSTRACT

We analyse, by doing very simple calculaltions, the internal degree of freedom leading to the de Broglie frequency associated to a material particle, as well, the confinement of quarks provided both by the Cornell potential and by the MIT bag model. We propose that the driving forces behind these confining models could be originated in the fluctuations of the metric, namely the particle interacting self-gravitationally, when its mass fluctuates in position throught of a distance equal to the Planck's length.

In the last years, a great deal of efforts has been dedicated to the study of quark confinement. As example, nonperturbative QCD was studed [1] by using the dual Landau - Ginzburg theory, obtaining as a result the linear static quark potential which characterizes the quark confinenent, due to the dual Meissner effect. Quark confinement also has been described within the formalism of the MIT bag model [2,3,4].

As was pointed out by Lenz, Moniz and Thies [5], despite many efforts, a generally accepted analytical explanation or qualitative description of confinement within the framework of QCD is still missing. The only theoretical model of a confining theory, which has a well defined derivation from QCD, is provided by the strong coupling limit of the lattice QCD. Unfortunately, this strong coupling limit does not distinguish QED and QCD as far as confinement is concerned. These two theories supposedly develop their characteristic differences only in a phase transition, which as a function of the coupling constant is known to occur for QED [6] and for which no evidence is found in lattice QCD calculations.

Taking in account these argumentations, it appears that the use of an alternative approach to treat the problem may be justifiable. So in this letter, we would like to show through some naïve considerations, that quark confinement could be related to the fluctuations of the space metric [7].

The starting point of this work is the Klein-Gordon (K-G) equation for a free particle. Paul [8] has used this equation in order to remove a possible ambiguity in the definition of the de Broglie frequency[9]. Let us write the K-G equation

$$\nabla^2 \psi - \frac{1}{c^2} \frac{\partial^2 \psi}{\partial t^2} = \left(\frac{mc}{\hbar}\right)^2 \psi, \qquad (1)$$

where m is the particle rest mass, c the speed of light and $\hbar$ the reduced Planck's constant.

Now let us consider only the time dependence of equation (1), namely

$$\frac{\partial^2 \psi}{\partial t^2} = -\omega_{dB}^2 \psi \ , \tag{2}$$

where $\omega_{dB} = \frac{mc^2}{\hbar}$ is the de Broglie frequency. If we multiply both sides of (2) by A, where A has a dimension of a characteristic length, we obtain

$$\frac{\partial^2 Y}{\partial t^2} = -\omega_{dB}^2 Y \ , \tag{3}$$

Where $Y = A\psi$ could be thought as a harmonic oscillator of amplitude A. A solution of (3) is given by

$$Y = A \sin \omega_{dB} t \ , \tag{4}$$

and by imposing the requirement that the amplitude of the velocity is equal to c we can write

$$\omega_{dB} A = \frac{mc^2}{\hbar} A = c \tag{5}$$

Equation (5) leads to

$$A = \frac{\hbar}{mc} \tag{6}$$

which identifies the characteristic length A with the reduced compton wavelength of the particle. We also notice that the maximum of $\partial^2 Y / \partial t^2$ could be thought as the amplitude $a_1$ of the aceleration, namely

$$a_1 = \omega_{dB}^2 A = \frac{mc^3}{\hbar} \ . \tag{7}$$

By multiplying $a_1$ (given by (7)) by m, we obtain $F_1$, the ampliude of the force acting on the particle. We have

$$F_1 = \frac{m^2 c^3}{\hbar} = \frac{Gm^2}{\lambda_P^2} \ , \tag{8}$$

where $\lambda_P^2 = \left(\dfrac{\hbar G}{c^3}\right)$ is the Planck's length and G is the gravitational constant. This amplitude of force can be interpreted as the particle self-gravitational interaction, when its mass fluctuates in position through a distance equal to the Planck's length.

Turning to the quark confinement problem let us write the following effective potential.

$$U = -\dfrac{\alpha_s \hbar c}{r} + Kr \ . \tag{9}$$

In the above potential, we have considered the interaction between a quark pair where r is the relative coordinate, $\alpha_s$ is the strong coupling constant with Kr (where K is the elastic constant) being the term responsible for the quark confinement. The above potential is sometimes refered in literature as the Cornell potential [1,10]. A discussion about quark confining potentials in relativistic equations is given by Ram [11].

Now, following Paul's procedure [8], we can write the following relation for the change in frequency

$$\hbar \delta\omega = -\delta U = -\left(\dfrac{\alpha_s \hbar c}{r} + Kr\right)\left(\dfrac{\delta r}{r}\right). \tag{10}$$

Dividing (10) by $2m_q c^2$, the "rest" energy of a quark pair and by using the definition of the de Broglie energy related to the quark constituent mass $m_q$, we can write

$$\left|\dfrac{\delta\omega}{2\omega_{dB}}\right| = \dfrac{1}{2m_q c^2}\left(\dfrac{\alpha_s \hbar c}{r} + Kr\right)\left(\dfrac{\delta r}{r}\right). \tag{11}$$

Making the requirement of "maximum fluctuability" by imposing that

$$\left|\dfrac{\delta\omega}{2\omega_{db}}\right| = \left(\dfrac{\delta r}{r}\right) = 1, \tag{12}$$

we get

$$Kr^2 - 2m_q c^2 r + \alpha_s \hbar c = 0. \tag{13}$$

Solving the above equation for r, we obtain

$$r = \frac{2m_q c^2 \pm \left(4m_q^2 c^2 - 4K\alpha_s \hbar c\right)^{1/2}}{2K} \quad . \tag{14}$$

When the term inside the radical of (14) vanishes, we have a threshold condition given by

$$KR_o = m_q c^2 \tag{15A}$$

and

$$m_q^2 c^4 = K\alpha_s \hbar c \; , \tag{15B}$$

where $R_0$ is the threshold value of r.

Equations (15A,B) imply that

$$\alpha_s \hbar c = kR_0^2 \tag{16}$$

which leads to the vanishing of U(r= $R_0$) (see(9)). From (15A) and (16) we also have

$$R_0 = \frac{\alpha_s \hbar}{m_q c}. \tag{17}$$

We observe that $R_0$ is not the nucleon radius, but a value of r where the two contributions to the Cornell potential both have the same absolute value. Besides this, we obtain from (15B) that

$$K = \frac{m_q^2 c^3}{\alpha_s \hbar} = \left(\frac{Gm_q^2}{\lambda_P^2}\right) \frac{1}{\alpha_s}. \tag{18}$$

Putting $\alpha_s = 1$, we see that K is equal to $F_1$, at $m = m_q$ (see(8)). Then the "string constant" K can be thought as a consequence of the fluctuations of the metric, as has occurred in the interpretation of "the driving force" leading to the de Broglie frequency. To make some estimate we put figures in the relation for $F_1(m_q)$, obtaining

$$K = F_1(m_q) = \frac{Gm_q^2}{\lambda_P^2} \cong 0.5 GeV/fm. \tag{19}$$

Values of the elastic constant of the linear contribution to the Cornell potential are generally estimated in the literature [4] as of order of 1 GeV/fm. This can be understood, if we consider that in the nucleon each quark interacts with the two other ones. So, we must multiply the number obtained in (19) by two, in order to account for the figures quoted in [4].

Now, let us look at another model which complies quark confinement. So, in the following, we are going to consider the MIT bag model [2,3,4,12]. As was pointed out by Brown and Rho [12], in the MIT bag model quarks are confined by fiat - by a boundary condition applied to the quark wavefunction at the radius R, the edge of of the "bag". Applying this boundary condition, one can verify that the normal component of the vector current is zero at r=R. Thus, no particles can escape from the bag. To allow collored quarks to exist locally, we must create a bubble, or bag, and this costs energy. The amount of energy is taken to be proportional to the volume: $\Delta E = \frac{4}{3}\pi R^3 B$, where B is the "bag constant". In order to obtain the quark ground-state energy, we proceed in an alternative way to that followed by Brown and Rho [12]. First we consider that each quark in the nucleon interacts with the other ones through the potential energy $E_s$ given by

$$E_s = -\frac{2\alpha_s \hbar c}{R} = -\frac{2\hbar c}{R}, \tag{20}$$

where $\alpha_s = 1$. Supposing that the negative strong coupling energy $E_s$ of the quark is just sufficient to cancel the positive masss energy of $m_q c^2$, so that the net energy of the quark is zero [13], we can write

$$E_q = m_q c^2 = \frac{2\hbar c}{R}, \tag{21}$$

and

$$R = \frac{2\hbar}{m_q c}. \tag{22}$$

Second, we write the energy of the bag with three quarks as [12]

$$E_{bag} = \frac{4}{3}\pi R^3 B + 3\left(\frac{2\hbar c}{R}\right). \tag{23}$$

To pursue further we notice that in the MIT bag model [2] the volume term $\Delta E$ corresponds to 1/4 of the nucleon rest energy. Then we can write

$$B \times \frac{4}{3}\pi R^3 = \frac{3}{4} m_q c^2, \tag{24}$$

where we have used the fact that the quark constituent mass is approximately one third of the nucleon mass. Putting (22) into (24) and solving for B, we obtain

$$B = \frac{9}{128\pi} \frac{m_q^4 c^5}{\hbar^3} = \left(\frac{9}{128\pi}\right) \frac{1}{\hbar c} \left(\frac{G m_q^2}{\lambda_P^2}\right)^2. \tag{25}$$

Therefore we verify that the bag constant B can be written as a constant which depends on $\hbar$ and c times the square of the force amplititude $F_1(m_q)$, where $F_1$ could be thought as the quark self-gravitational interaction when its constituent mass fluctuates in position through of a distance equal to the Planck's length. As $F_1$ appears squared in the relation defining B, we can interpret the bag constant as due to a kind of van der Walls interaction among the quarks which constitute the nucleon. Putting numbers in (25), we obtain that

$B \cong 8 \times 10^{28} \, atm$,

which is the same as the figure quoted by Jaffe [3], on estimating the bag constant value.

By concluding, it seems that quark cofinement can be interpreted as a result of the gravitational interaction of a particle with its "ghost" partner, the two masses being separeted by a "fluctuating" distance equal to the Planck's length. The same kind of driving force appears to be responsible for the "internal motion" which sets up the clock originated from the association of the de Broglie frequency to a material particle [14,15,16,17].